\documentclass[trackchanges,twocolumn]{aastex701}
\usepackage{amsmath,amstext}
\usepackage[T1]{fontenc}
\usepackage{multirow}
\usepackage[utf8]{inputenc} 
\usepackage{float}
\usepackage{graphicx}

\usepackage{tabularx} 
\usepackage{booktabs}

\shorttitle{TTVs of TOI-2109 b}
\shortauthors{Greklek-McKeon et al.}

\begin{document}

\title{JWST and HST Confirm Transit Timing Variations for the Ultra-Hot Jupiter TOI-2109 b}

\correspondingauthor{Michael Greklek-McKeon}
\email{mgreklek-mckeon@carnegiescience.edu}

\author[0000-0002-0371-1647]{Michael Greklek-McKeon}
\affiliation{Earth and Planets Laboratory, Carnegie Institution for Science, 5241 Broad Branch Road, NW, Washington, DC 20015, USA}
\email{mgreklek-mckeon@carnegiescience.edu}

\author[0000-0001-9164-7966]{Julie Inglis}
\affiliation{Department of Astronomy \& Astrophysics, University of California San Diego, 9500 Gilman Drive, La Jolla,  CA 92093, USA}
\email{jinglis@ucsd.edu}

\author[0000-0003-0354-0187]{Nicole L. Wallack}
\affiliation{Earth and Planets Laboratory, Carnegie Institution for Science, 5241 Broad Branch Road, NW, Washington, DC 20015, USA}
\email{nwallack@carnegiescience.edu}

\author[0000-0003-2527-1475]{Shreyas Vissapragada}
\affiliation{Carnegie Science Observatories, 813 Santa Barbara Street, Pasadena, CA 91101, USA}
\email{svissapragada@carnegiescience.edu}

\author[0000-0001-9665-8429]{Ian Wong}
\affiliation{Space Telescope Science Institute, 3700 San Martin Drive, Baltimore, MD 21218, USA}
\email{iwong@stsci.edu}

\author[0009-0006-1653-3524]{John R. T. Allen}
\affiliation{Department of Physics (Atmospheric, Oceanic, and Planetary Physics), University of Oxford, Oxford OX1 3PU, UK}
\email{john.allen@physics.ox.ac.uk}

\author[0000-0003-1728-8269]{Yayaati Chachan}
\affiliation{Department of Astronomy \& Astrophysics, University of California, Santa Cruz, CA 95064, USA}
\email{ychachan@ucsc.edu}

\author[0000-0002-8958-0683]{Fei Dai}
\affiliation{Institute for Astronomy, University of Hawai`i, 2680 Woodlawn Drive, Honolulu, HI 96822, USA}
\email{fdai@hawaii.edu}

\author[0000-0002-3263-2251]{Guangwei Fu}
\affiliation{Department of Physics and Astronomy, Johns Hopkins University, Baltimore, MD, USA}
\email{guangweifu@gmail.com}

\author[0000-0003-3759-9080]{Tiffany Kataria}
\affiliation{Jet Propulsion Laboratory, California Institute of Technology, 4800 Oak Grove Drive, Pasadena, CA 91109, USA}
\email{tiffany.kataria@jpl.nasa.gov}

\author[0000-0002-9258-5311]{Thaddeus D. Komacek}
\affiliation{Department of Physics (Atmospheric, Oceanic, and Planetary Physics), University of Oxford, Oxford OX1 3PU, UK}
\email{tad.komacek@physics.ox.ac.uk}

\author[0009-0008-2801-5040]{Johanna K. Teske}
\affiliation{Earth and Planets Laboratory, Carnegie Institution for Science, 5241 Broad Branch Road, NW, Washington, DC 20015, USA} 
\affiliation{The Observatories of the Carnegie Institution for Science, 813 Santa Barbara St., Pasadena, CA 91101, USA}
\email{jteske@carnegiescience.edu}

\begin{abstract}
 
Ultra-hot Jupiters are a rare yet valuable subset of exoplanets that can help elucidate both the internal properties of their host stars as well as the orbital evolution of close-in planets. An extreme among extremes, TOI-2109~b is a super-Jupiter ($1.347 \pm 0.047 R_{\rm Jup}$, $5.02 \pm 0.75 M_{\rm Jup}$) with the shortest orbital period (0.67247479 $\pm$ 0.00000028 days) and the fastest predicted orbital decay rate of any known giant planet. Previous observations hinted at a quadratic long-term trend in the transit timing variations (TTVs) consistent with orbital decay, as well as short-term sinusoidal TTVs consistent with a near-resonant planetary companion that would indicate a dynamically quiescent migration pathway. Here, we combine archival ground-based, TESS, and CHEOPS photometry with new transit and eclipse data from HST and JWST to update the TTV analysis. We find very strong evidence for sinusoidal TTVs over a linear ephemeris ($\Delta$BIC of 33.6), consistent with a nearby planetary companion as suggested by \cite{Harre2024}, and rule out several non-planetary scenarios that could explain the sinusoidal TTVs. We also find strong evidence of a quadratic TTV trend ($\Delta$BIC of 6.75) consistent with an orbital decay rate of -3.46 $\pm$ 1.17 ms/yr and a stellar tidal quality factor of 2.13$^{+1.18}_{-0.61} \times10^7$, though we cannot rule out line-of-sight acceleration as the cause of the quadratic trend. If our results are confirmed, TOI-2109~b would be the only known exoplanet besides WASP-12~b with a decaying orbit and the most extreme of an ultra-rare class of close-in giant planets with nearby companions.

\end{abstract}


\section{Introduction} \label{sec:style}

Ultra-hot Jupiters are highly irradiated giant planets on extremely short period orbits with dayside temperatures $\gtrsim$ 2000 K \citep[e.g.,][]{Gaudi2017}. These planets represent a tiny fraction of the overall exoplanet population \citep[$< 1\%$, NASA Exoplanet Archive 2026,][]{Christiansen2025}, but the extreme tidal, radiative, and dynamical processes affecting ultra-hot Jupiters makes them uniquely valuable for studying planet formation and evolution. The fate of close-in giant planets, including whether tidally-driven orbital decay causes them to spiral into their host stars, is set by the efficiency of tidal dissipation within the star. This efficiency is parameterized by the modified stellar tidal quality factor $Q_\star'$ \citep{Goldreich1996}. Theoretical predictions for $Q_\star'$ span several orders of magnitude, depending on the dominant dissipation mechanism, but values in the range of $Q_\star' \approx 10^{6}$ to $10^{8}$ are typically expected for F-type stars \citep{Ogilvie2007, Lanza2011}. This uncertainty translates directly into large uncertainties on predicted orbital decay timescales, and it remains an open question what fraction of the shortest period giant planets will survive the main sequence lifetimes of their hosts \citep[e.g.,][]{Adams2024}. 

While many ultra-hot Jupiters have predicted decay rates detectable on timescales of years to decades \citep[e.g.,][]{Adams2024}, only one has a confirmed detection. WASP-12 b remains the only unambiguous detection of tidally driven orbital decay, with an orbit shrinking at $\sim$30\,ms\,yr$^{-1}$ \citep{Maciejewski2016, Patra2017, Yee2020,Turner2021,Wong2022,Nediyedath2023,Kutluay2023,Leonardi2024,Shen2024,Kutluay2026,Biswas2026}. Beyond WASP-12, the many dedicated searches for orbital decay across the close-in giant planet population have yielded only marginal detections, upper limits, and non-detections \citep[e.g.,][]{Ivshina2022, Adams2024, Kutluay2026}. This highlights how both rare and observationally challenging these confirmations of orbital decay are, and how useful a new detection would be for constraining the physics of stellar interiors and understanding the fate of ultra-hot Jupiters.

TOI-2109~b is a super-Jupiter ($1.347 \pm 0.047 R_{\rm Jup}$, $5.02 \pm 0.75 M_{\rm Jup}$) on the shortest orbital period ($0.67247479 \pm 0.00000028$~days) of any known giant planet. TOI-2109 is an F-type star with a mass, radius, and effective temperature of $1.453 \pm 0.074 M_\odot$, $1.698 \pm 0.062 R_\odot$, $T_{\rm eff} = 6540 \pm 160$ K, respectively. The system was discovered and characterized by \cite{Wong2021}, who identified it as one of the hottest-known planets. Given its extreme proximity to the host star, TOI-2109~b was identified as the single most promising candidate for the detection of tidal orbital decay. \cite{Wong2021} estimated that for stellar tidal quality factors in the range $Q_\star' = 10^{5}$--$10^{7}$, the orbital decay rate of TOI-2109~b would be $\sim$10--740\,ms\,yr$^{-1}$, corresponding to an inspiral timescale as short as a few Myr for the most efficient dissipation. For reasonable values of $Q_\star'$, decay-induced transit timing variations (TTVs) on the order of minutes should therefore have accumulated over the observational baseline that is now available. The system was subsequently monitored with CHEOPS by \cite{Harre2024}, who identified a crucial precovery transit in archival WASP data that extends the timing baseline to more than a decade, and it was revisited with additional TESS and ground-based data by \cite{Alvarado2025}. Both of these more recent studies searched for the expected decay signal and placed increasingly stringent lower limits on $Q_\star'$, but with best-fit decay rates that were consistent with a constant period at the $\sim$2-3$\sigma$ level, depending on the treatment of sinusoidal TTVs that may also be present in the data as suggested by \cite{Harre2024}. 

How ultra-hot Jupiters arrive at their observed locations is also an open question, with proposed pathways including disk-driven migration, high-eccentricity migration, and in-situ formation \citep[e.g.,][]{Dawson2018}. These pathways yield different predictions for the presence of nearby planetary companions. High-eccentricity migration is expected to disrupt or eject most nearby planetary companions, while the relatively dynamically quiet pathways of disk migration and in-situ formation could leave planetary companions undisturbed. The recent discoveries of close companions to a small number of hot Jupiters have therefore revealed that at least some fraction of these planets must have reached their present orbits through a dynamically quiescent pathway. These systems include WASP-47 \citep{Becker2015,Nascimbeni2023}, Kepler-730 \citep{Canas2019}, TOI-1130 \citep{Huang2020,Korth2023}, TOI-2000 \citep{Sha2023}, WASP-84 \citep{Maciejewski2023}, TOI-1408 \citep{Korth2024}, WASP-132 \citep{Hord2022,Grieves2025}, TOI-5143 \citep{Quinn2026}, TOI-4468 \citep{Livesey2026}, and the candidate system TOI-2818 \citep{McKee2025}. Population-level analyses indicate that the total fraction of hot Jupiters with nearby companions is low, around $\sim$ 10\% \citep[e.g.,][]{Hord2021,Wu2023,Sha2026}, setting a lower limit on what fraction of hot Jupiters either form in situ or undergo disk migration.

\cite{Harre2024}, hereafter referred to as JVH24, reported evidence for a low-amplitude sinusoidal TTV signal in the CHEOPS transit times of TOI-2109~b, which they attributed to an external, non-transiting, near-resonant planetary companion. If confirmed, such a companion to the shortest orbital period giant planet would be a valuable test of the extreme limits on companion stability that define which planet formation and evolution pathways are possible for ultra-hot Jupiters. The inner edges of protoplanetary disks can be just a few stellar radii away from the host star \citep[e.g.,][]{Dullemond2010}, but TOI-2109~b's extreme proximity to TOI-1209 \citep[$a$ = 0.01791 $\pm$ 0.00065 au, $a/R_{\star}$ = 2.268 $\pm$ 0.021,][]{Wong2021} implies that if it formed in-situ, it must have been at the extreme inner edge of the disk. Previous theoretical, demographics, and modeling work has suggested that hot Jupiters can form in situ near the inner edges of their protoplanetary disks and be accompanied by low-mass planetary companions \cite[e.g.,][]{Batygin2016,Bailey2018,Mathur2025}. This is relevant for the purported sinusoidal trend in the TTVs consistent with a nearby planetary companion from JVH24, but the proximity ($< 0.02$ au) and high mass ($5.02 \pm 0.75 M_{\rm Jup}$) of TOI-2109~b are beyond the limits of typical models for hot Jupiter in situ formation.

Alternatively, TOI-2109~b may have formed further out and reached its present day orbit through disk migration or high-eccentricity tidal migration. In the disk migration scenario, the planet migrates inward through torques from the gaseous disk \citep[e.g.,][]{Goldreich1980,Lin1986,Baruteau2014}, eventually reaching its short orbital period when the disk dissipates, or more likely in the case of TOI-2109~b, when the orbital migration is halted by a magnetocavity at the innermost disk edge \citep[e.g.,][]{Rice2008,Chang2010}. In the high-eccentricity migration scenario, the planet is initially perturbed onto a highly elliptical orbit through a dynamical interaction such as planet-planet scattering \citep[e.g.,][]{Rasio1996,Weidenschilling1996}, cyclic secular interactions \citep[e.g.,][]{Kozai1962,Lidov1962,Wu2003,Fabrycky2007}, or chaotic secular interactions \citep[e.g.,][]{Wu2011}, followed by a process where tides raised by the star on the planet work to shrink and circularize the orbit \citep[e.g.,][]{Eggleton1998}. However, high-eccentricity migration is dynamically violent and often leads to orbital disruption for nearby companion planets (e.g., \cite{Mustill2015,Sha2026}, see \cite{Fortney2021} for further review). Therefore, if there is indeed a nearby planetary companion to TOI-2109~b, it would indicate that the shortest orbital period ultra-hot Jupiter likely reached its current orbit through disk migration. Independent verification of both the sinusoidal TTV and quadratic orbital decay TTV trends are therefore especially important.
%In this work we are able to provide strong evidence for both trends, albeit with certain caveats requiring further modeling and observations to be completely certain that the causes of these sinusoidal and quadratic TTV signals are a nearby planetary companion and orbital decay.

In this work, we combine the full set of archival ground-based, TESS, and CHEOPS photometry used in previous dynamical analyses of TOI-2109~b \citep{Harre2024,Alvarado2025} with new high precision transit and eclipse observations from JWST (GO 8877, PIs N. Wallack \& J. Inglis) and HST (GO 17543, PI G. Fu) to improve the dynamical constraints on TOI-2109~b. The extreme timing precision of the new JWST data, at the $\sim$3 second level, anchors the transit timing baseline and enables us to drastically improve the dynamical constraints, allowing us to firmly detect both the sinusoidal TTV signal and a quadratic TTV trend consistent with orbital decay. We describe our observations and timing measurements in Section~\ref{sec:Observations}, present our dynamical analysis of the sinusoidal and quadratic TTV signals and our assessment of alternative dynamical explanations in Section~\ref{sec:Dynamical_Analysis}, and summarize our conclusions in Section~\ref{sec:Conclusions}.

\section{Observations and Timing Measurements} \label{sec:Observations}

To update the dynamical constraints on TOI-2109~b, we combined new space-based observations from JWST and HST with the full set of archival photometry that has been used in previous dynamical analyses of the system \citep{Wong2021,Harre2024,Alvarado2025, Kutluay2026}. In the following subsections, we describe the origin and summarize the data reduction procedures for the archival ground-based (\ref{subsec:obs_ground}), CHEOPS (\ref{subsec:obs_CHEOPS}), and TESS (\ref{subsec:obs_TESS}) data, followed by the new HST (\ref{subsec:obs_HST}) and JWST (\ref{subsec:obs_JST}) observations.

\subsection{Ground-based Photometry} \label{subsec:obs_ground}

TOI-2109~b was observed from the ground by 11 facilities across 20 full and partial transit observations as part of the initial planet discovery and validation effort described in \cite{Wong2021}. These observations were taken as part of the TESS Follow-up Observing Program (TFOP) Sub Group 1 (SG1; seeing-limited time-series photometry) collaboration. The original ground-based TFOP SG1 observations published in \cite{Wong2021} were taken between July 2020 and June 2021. These observations span a range of photometric bandpasses (from $B$ to $z_s$), image resolutions, exposure times (3-100 seconds), and photometric precisions. \cite{Wong2021} chose not to include partial transits in joint fitting of the ground-based data due to the possibility of significant biases to the transit timing and transit shape parameters, so we also do not include any measured times from partial ground-based transit events in our analysis. 

After \cite{Wong2021}, four additional TFOP SG1 observations were made between June 2021 and August 2023. These observations were analyzed in \cite{Alvarado2025}, hereafter referred to as JAM25, which also independently analyzed the previous 20 transit observations originally published in \cite{Wong2021}. In brief, JAM25 analyzed the available ground-based transit data with the \texttt{PyTransit} package \citep{Parviainen2015} using a \cite{Mandel2002} transit model by first phase-folding all transit events to find reliable values for the fitted transit shape parameters (planet-to star radius ratio, stellar density, impact parameter, orbital period, zero epoch, eccentricity, and argument of periastron), followed by fixing these values when fitting each individual transit to recover the transit times. After processing the ground-based transit data, JAM25 discarded any times with uncertainties larger than 2 minutes, since these data are uninformative for the small TTV amplitudes relevant for TOI-2109~b. We adopt the same set of ground-based transit times for our analysis.

In addition to the SG1 photometry, TOI-2109 was observed by several cameras of SuperWASP-N \citep[Wide Angle Search for Planets, ][]{Pollacco2006} from 2006 to 2011. JVH24 phase-folded and analyzed this data, making a key precovery detection of the transit of TOI-2109~b that enables long-baseline searches for orbital decay (see Section \ref{subsec:Decay}). Rather than re-reducing these ground-based data, we adopt the reference transit times of JAM25 for SG1 sources, and JVH24 for the WASP data. All TFOP SG1 observations used for ground-based transit times in this paper are publicly available through the Exoplanet Follow-up Observing Program (ExoFOP)\footnote{\url{https://exofop.ipac.caltech.edu/tess/}}. Detailed descriptions of the ground-based data reductions are available in \cite{Wong2021} and JAM25, and the final set of ground-based transit timing measurements from JAM25 is available on GitHub\footnote{\label{github_note}\url{https://github.com/JAAlvarado-Montes/TOI-2109b}}.

\subsection{CHEOPS} \label{subsec:obs_CHEOPS}

TOI-2109~b was observed by the CHaracterising ExOPlanet Satellite \citep[CHEOPS, ][]{Benz2021} over 31 visits from 2021 to 2023, with 23 transit observations and 8 phase curve observations as detailed in JVH24. All of these are publicly accessible via the CHEOPS archive\footnote{\url{https://cheops-archive.astro.unige.ch/archive_browser/}}. JVH24 performed two reductions of the CHEOPS data, one using the standard CHEOPS Data Reduction Pipeline (DRP), and the other by performing aperture photometry on the imagette subarrays with PSF Imagette Photometric Extraction \citep[PIPE,][]{Brandeker2024}. JVH24 derived individual transit times for both the DRP and PIPE reductions, and found inconsistencies for some partial transit observations. These biased transit times were filtered out, leaving 26 total CHEOPS transit times that agreed within 1$\sigma$ between both data reductions. We adopt these CHEOPS PIPE transit times from JVH24 in our analysis. We note that these CHEOPS observations and their associated analysis in JVH24 are of particular importance to this study, as they were responsible for first identifying the low amplitude sinusoidal TTV signature that we confirm in Section \ref{subsec:Sinusoidal_TTVs}.

\begin{deluxetable}{cccc}
\tabletypesize{\footnotesize}
\tablecaption{Observed transit times of TOI-2109~b.\label{tab:transit times}}
\tablehead{
  \colhead{$T_i$ (BJD$_{\rm TDB}$ - 2457000)} &
  \colhead{$\sigma_{T_i}$ (days)} &
  \colhead{Facility} &
  \colhead{Reference}
}
\startdata
2453885.68861800 & 0.001079000 & WASP & JVH24 \\
2458984.38956203 & 0.000715023 & TESS & JAM25 \\
2458985.06194799 & 0.000800749 & TESS & JAM25 \\
2458985.73373064 & 0.000576356 & TESS & JAM25 \\
\enddata
\tablecomments{Best-fit transit times ($T_i$) and their $1\sigma$ uncertainties used in the dynamical analysis of this paper. This table is shown in an abbreviated form to illustrate its form and content, and the table in its entirety is available in machine-readable form online.}
\end{deluxetable}

\subsection{TESS} \label{subsec:obs_TESS}

TOI-2109 was observed by the Transiting Exoplanet Survey Satellite \citep[TESS, ][]{Ricker2014} in 3 sectors (25, 52, and 79) in 2020, 2022, and 2024, with cadences of 30, 2, and 0.33 minutes, respectively. The sector 25 data were originally analyzed in \cite{Wong2021}, while the new sectors 52 and 79 data were analyzed together with sector 25 in JAM25. All TESS data are publicly available via the Mikulski Archive for Space Telescopes (MAST)\footnote{\url{https://mast.stsci.edu/}}. JAM25 processed the TESS light curves and measured transit times for individual events using the formalism and code for transit timing analysis presented in \cite{Ivshina2022}, with additional details available in JAM25. The resulting transit times have highly variable precisions, due to the photometric quality across TESS sectors. Some sector 25 transit times have uncertainties close to 10 minutes, so JAM25 performs the same data clipping procedure as for the ground-based transit data, where any event with a timing uncertainty larger than 2 minutes is clipped. We adopt these TESS transit times in our analysis, which resulted in 26, 21, and 21 transit times from sectors 23, 52, and 79, respectively. These measured TESS transit times from JAM25 are also available on GitHub. %\footnote{\url{https://github.com/JAAlvarado-Montes/TOI-2109b}}.

\subsection{JWST Observations and Data Reduction} \label{subsec:obs_JST}

We observed one full orbital phase curve including two eclispes of TOI-2109~b using NIRSpec with the G395H/F290LP grating, SUB2048 subarray, and 26 groups per integration beginning on 14 August 2025 (JWST GO 8877, PIs N. Wallack \& J. Inglis). The spectral trace was broken up over two simultaneously utilized detectors, NRS1 and NRS2, covering 2.87--5.18 $\mu$m with a gap in wavelength coverage between 3.72 and 3.82 $\mu$m. The atmospheric constraints and a thorough detailing of the data analysis methods for the JWST observations will be presented in upcoming papers, therefore we only briefly describe the data analysis methods herein.

We analyze the JWST observation using \texttt{Eureka!}, an end-to-end pipeline for the analysis of HST and JWST observations \citep{Bell2022}. When analyzing JWST observations, \texttt{Eureka!} acts as a wrapper around the \texttt{jwst} pipeline. We use version 1.3 of \texttt{Eureka!} and  version 1.20.2 of the \texttt{jwst} pipeline with CRDS context pmap 1364. When reducing the observation, we utilize the default Stage 1 and Stage 2 parameters with the exception of using 15$\sigma$ for the jump rejection threshold. We also utilize the group-level background subtraction within \texttt{Eureka!} to correct for the 1/$f$ noise. In Stage 3, we extract the spectrum using apertures of 4 pixel half-widths and 8 pixel background half-widths. Herein we only utilize the NRS1 and NRS2 extracted white light curves from Stage 4, with no further division by wavelength.

\subsection{Phase Curve Fitting} \label{subsec:phase}

In order to derive transit and secondary eclipse times for the JWST and HST observations, we fit the eclipses simultaneously with the planetary phase curve. The total model for our white light curves for NRS1, NRS2, and HST is a combination of three terms:
\begin{equation}
    F(t) = S(t)\times(T(t)\times F_*(t)+F_p(t))
\end{equation}
where $F_*(t)$ is the baseline stellar flux including the contributions from effects such as stellar Doppler beaming, stellar ellipsoidal distortion and stellar variability, T(t) is the planet transit model generated using the \texttt{batman} package \citep{batman}, $F_p$(t) is the flux contribution from the planet, and S(t) are the respective instrument systematic models. The planetary flux contribution is modeled as: 
\begin{equation}
    F_p(t) = (E(t)-1)\times \Phi(\phi),
\end{equation}
where $E(t)$ is the secondary eclipse model generated using \texttt{batman} \citep{batman}, and $\Phi(t)$ is the phase-dependent planet flux. We find that a second degree sinusoid model is required to capture the planet flux variation as a function of orbital phase  :
\begin{multline}
    \Phi(\phi) = 1+A (\cos{\phi}-1) + B \sin{\phi} \\ + C (\cos{2\phi}-1) + D \sin{2\phi}, \label{eqn:pc}
\end{multline}
where $\phi$ = $2\pi(t-T_e)/P$. Higher order terms are disfavored by the data. We fix the limb darkening coefficients to values generated using the package \texttt{ExoTiC-LD} \citep{Grant2022}. For our JWST observations, we fit for a free transit and two secondary eclipse midpoint times. For our HST observations, we instead derive a single transit and secondary eclipse time from jointly fitting the phase curve. As our two HST phase curve observations are obtained one year apart, we find that including an additional sinusoidal function at a period of 0.6 days is required to jointly model the planetary phase curve. However, fits with and without the stellar variability term included change the derived transit and eclipse times by $<1\sigma$, suggesting this effect does not bias the derived timings.  

More details on the data reduction and phase curve fits can be found in the upcoming companion papers. The resulting best-fit transit time from the JWST phase curve is reported in Table \ref{tab:transit times}.

\subsection{HST Observations and Data Reduction} \label{subsec:obs_HST}

We also observed two full orbital phase curves of TOI-2109~b with HST/WFC3 G141 (HST GO 17543, PI Fu). Visit 1 began on 27 March 2024 and visit 2 began on 28 May 2025. Each of the two visits consists of 15 orbits and includes two secondary eclipses and one transit with two orbits occurring prior to the first eclipse. As the atmospheric constraints from these observations will also be presented in forthcoming papers where the data reduction details will be thoroughly detailed, we only briefly describe the analysis herein. 

We reduce the HST observations using the open-source package \texttt{PACMAN} \citep{Zieba2022}. Beginning with the \texttt{.ima} files obtained from the Barbara A. Mikulski Archive for Space Telescopes (MAST), we then follow the default data reduction steps within \texttt{PACMAN}. We correct the timestamps of the \texttt{.ima} files to account for the motion of HST with respect to the barycenter of the Solar System. We perform wavelength calibration using the \cite{castelli_new_2003} stellar model grid with the best-fit parameters for TOI-2109 \citep{Wong2021}. We mask bad pixels that have been flagged by calwf3 with data quality DQ = 4 or 512. We perform background subtraction by taking the median of pixels with a flux value under 1000 excluding a window of 20 pixels centered on the trace. We then extract the white light curves from the two visits using the optimal extraction algorithm from \cite{Horne1986}, using  an aperture size of 12. We use the same phase curve model as described in Section \ref{subsec:phase} with the transit time as a free parameter to measure the best-fit midtime. Unfortunately one of our HST phase curves did not sample any data during transit ingress or ingress due to Earth occultations, and therefore the transit timing measurement precision is not useful for TTV analysis. The other transit time from HST is reported in Table \ref{tab:transit times}.

\section{Dynamical Analysis} \label{sec:Dynamical_Analysis}

After assembling the full set of archival and new transit times (Section \ref{sec:Observations}), we used them to characterize the dynamical state of the TOI-2109 system. We first investigated the presence of low-amplitude sinusoidal TTVs as suggested by JVH24 to test whether the extreme precision of our new JWST transit combined with new TESS and ground-based data from JAM25 can confirm or refute this signal (Section \ref{subsec:Sinusoidal_TTVs}). After identifying the strong signature of sinusoidal TTVs consistent with originally reported TTV amplitude and period constraints from JVH24, we searched for evidence of tidally driven orbital decay by constraining the presence of a quadratic long-term TTV trend \citep[e.g.,][]{tidal_circ,Maciejewski2016,Yee2020} in addition to the short-term sinusoidal TTVs (Section \ref{subsec:Decay}). Finally, we analyze whether alternative dynamical configurations could produce sinusoidal or quadratic TTV signatures (Sections \ref{subsec:Other causes of sinusoidal TTVs}, \ref{subsec:LOS Acceleration}). The transit times used throughout this section, including the source facilities, observation dates, and original source publications are available in Table \ref{tab:transit times}.

\subsection{Constraints on Sinusoidal TTVs} \label{subsec:Sinusoidal_TTVs}

Evidence for sinusoidal TTVs indicative of a potential outer companion planet to TOI-2109~b was presented in JVH24, based on variations in the CHEOPS transit times relative to a linear ephemeris. JVH24 noted that this low-amplitude ($<$ 1 minute) sinusoidal trend is only present in the CHEOPS data due to the high precision (median timing $\sigma$ $\sim$ 20 seconds, maximum timing $\sigma$ $<$ 1 minute). JVH24 attribute this trend to a nearby, non-transiting planetary companion, and they modeled this signal with a sinusoid of the form:
\begin{equation} \label{eq:sinusoid}
t_{\rm tra}(N) = T_0 + NP + A\cos\!\left[\frac{2\pi}{P_{\rm sup}}\left(N\,P - T_{\rm sup}\right)\right]
\end{equation}
where $T_0$ is the reference mid-transit time, $P$ is the orbital period, N is an integer representing the orbit number, and $A$, $P_{\rm sup}$, and $T_{\rm sup}$ are the amplitude, super-period (i.e. the TTV oscillation period), and reference epoch of the sinusoidal TTV, respectively. JVH24 fit this model to their CHEOPS transit times and found that two super-periods, $P_{\rm sup} \approx 88$\,d and $P_{\rm sup} \approx 117$\,d, provided comparably good fits to the data. The 117 d sinusoid was preferred over a simple linear ephemeris by $\Delta\mathrm{BIC} = 4.0$, while the 88 d sinusoid only slightly lower at $\Delta\mathrm{BIC} = 3.86$.

Model comparisons throughout this work, as well as in JVH24, are performed using the Bayesian Information Criterion \citep[BIC;][]{Schwarz1978}, $\mathrm{BIC} = \chi^2 + k\ln n$, where $k$ is the number of free parameters and $n$ is the number of transit times being modeled. We interpret BIC differences following the guidelines of \cite{Raftery1995}, for which $2 < \Delta\mathrm{BIC} < 6$ represents positive evidence for one model over another, $6 < \Delta\mathrm{BIC} < 10$ is strong evidence, and $\Delta\mathrm{BIC} > 10$ is very strong evidence for the preferred (lower BIC) model. Therefore, the original sinusoid model for the TTVs in CHEOPS data from JVH24 represented positive but not yet strong evidence for the presence of sinusoidal TTVs.

JVH24 supported their interpretation of the sinusoidal TTVs as evidence for a nearby planetary companion by performing $N$-body modeling using the \texttt{TRADES} code \citep{Borsato2014, Borsato2019} and photo-dynamical modeling with \texttt{PyTTV} \citep{Korth2023}. This photo-dynamical model combines the REBOUND n-body modeling code \citep{Rein2012,Rein2015,Tamayo2020} with the \texttt{PyTransit} light curve modeling code  \citep{Parviainen2015,ParviainenKorth2020,Parviainen2020} to jointly fit the photometric light curves, radial velocity observations, and individual transit times. JVH24 notes that any planet coplanar but interior to TOI-2109~b should transit, but they found no additional transiting planet signatures in the TESS data, therefore any planetary companion causing TTVs is likely to be external to TOI-2109~b. The TTV modeling of JVH24 found that a near-resonant (orbital period ratio of 5:3, 2:1, 7:3, 5:2, 3:1, or 7:2) outer companion (period from 1.12 to 2.35 days) with a mass up to 2 $M_{Jup}$ can reproduce the observed sinusoidal TTV amplitude and period, though the orbital periods and masses of non-transiting companions from TTV solutions are inherently degenerate \citep{Lammers2026}. See JVH24 for additional details on the mapping of the allowable parameter space for a planetary companion causing these sinusoidal TTVs. 

In principle, radial velocity (RV) data could provide an additional constraint on the presence of any external planetary companion causing the sinusoidal TTVs, but due to the high rotational speed of the star, the RV precision is poor \citep{Wong2021}. The median precision of the RV measurements in \cite{Wong2021} is 410 m/s, with best-fit RV jitter terms (to account for additional noise in the data) in the 300 m/s range. Given the  masses ($< 350 M_{\oplus}$) and orbital periods ($<$ 2.4) of the external planetary companion constrained by the photo-dynamical analysis of JVH24, the expected RV semiamplitudes of planet c are $<$ $\sim$150 m/s, which means the existing RV data are uninformative for confirming the source of the sinusoidal TTV as a nearby planetary companion. Similarly, the expected astrometric signal from TOI-2109~b plus the most massive companion planet allowed by the analysis of JVH24 is $< 0.4$ $\mu$as, well below the detectability threshhold of Gaia \citep[e.g.,][]{Feng2026}.

\subsubsection{CHEOPS + JWST Sinusoidal TTV Analysis} \label{subsubsec:cheops + jwst ttvs}

To test whether this sinusoidal TTV signal is real, we compared Equation \ref{eq:sinusoid} with a linear ephemeris model, $t_{\rm tra}(N)$ =  $T_0 + NP$, where $T_0$ is the initial transit time, $P$ is the orbital period, $N$ is the transit epoch relative to the initial transit in the data being fit, and $t_{\rm tra}(N)$ is the transit time at the $N$-th epoch. We began by fitting the CHEOPS transit times from JVH24 and our extreme precision ($\sim$3 second uncertainty) JWST transit time from Table \ref{tab:transit times}. Each model was fit using weighted least squares via the \texttt{scipy} \texttt{curve\_fit} function \citep{Virtanen2020}, which utilizes the Levenberg–Marquardt algorithm for non-linear least squares fitting and uses the per-point transit time uncertainties as weights. 

Because there are large gaps in the observational baseline, the likelihood of the sinusoidal model is highly multimodal in $P_{\rm sup}$, so we do not optimize $P_{\rm sup}$ directly from an initial guess. Instead, we scan a dense grid of trial super-periods (10,000 points in log space from 10 days to twice the observational baseline), where we fit for the remaining parameters at each fixed super-period value and then compute the $\chi^2$ values, to identify which is the best-fitting super-period. We then adopt the grid minimum $P_{\rm sup}$ as the starting point for a full least-squares fit of all five parameters in the sinusoidal model. 

From the CHEOPS and JWST data alone, we found that a sinusoidal model with $P_{\rm sup} = 88.1$ days is preferred over the linear ephemeris with $\Delta\mathrm{BIC} = 5.9$, which by the \cite{Raftery1995} scale constitutes positive evidence. This is stronger evidence than originally found in JVH24 ($\Delta\mathrm{BIC} = 4.0$), though not yet a strong detection based on the scale of \cite{Raftery1995}. We note that another peak in the $P_{\rm sup}$ distribution at $\sim$ 176.7 days, close to twice the best-fit period, provides a similar quality fit to the CHEOPS + JWST data ($\Delta\mathrm{BIC} = 5.1$). JVH24 also identified an $\sim$ 88 day sinusoidal signal with only a marginally weaker significance than their best-fit $P_{\rm sup}$, indicating that our TTV amplitude and $P_{\rm sup}$ constraints are consistent with those of JVH24 (See our Figure \ref{fig:Sinusoid JWST periods} compared to Figure 3 of JVH24).

\begin{figure}
\begin{center}
  \includegraphics[width=8.5cm]{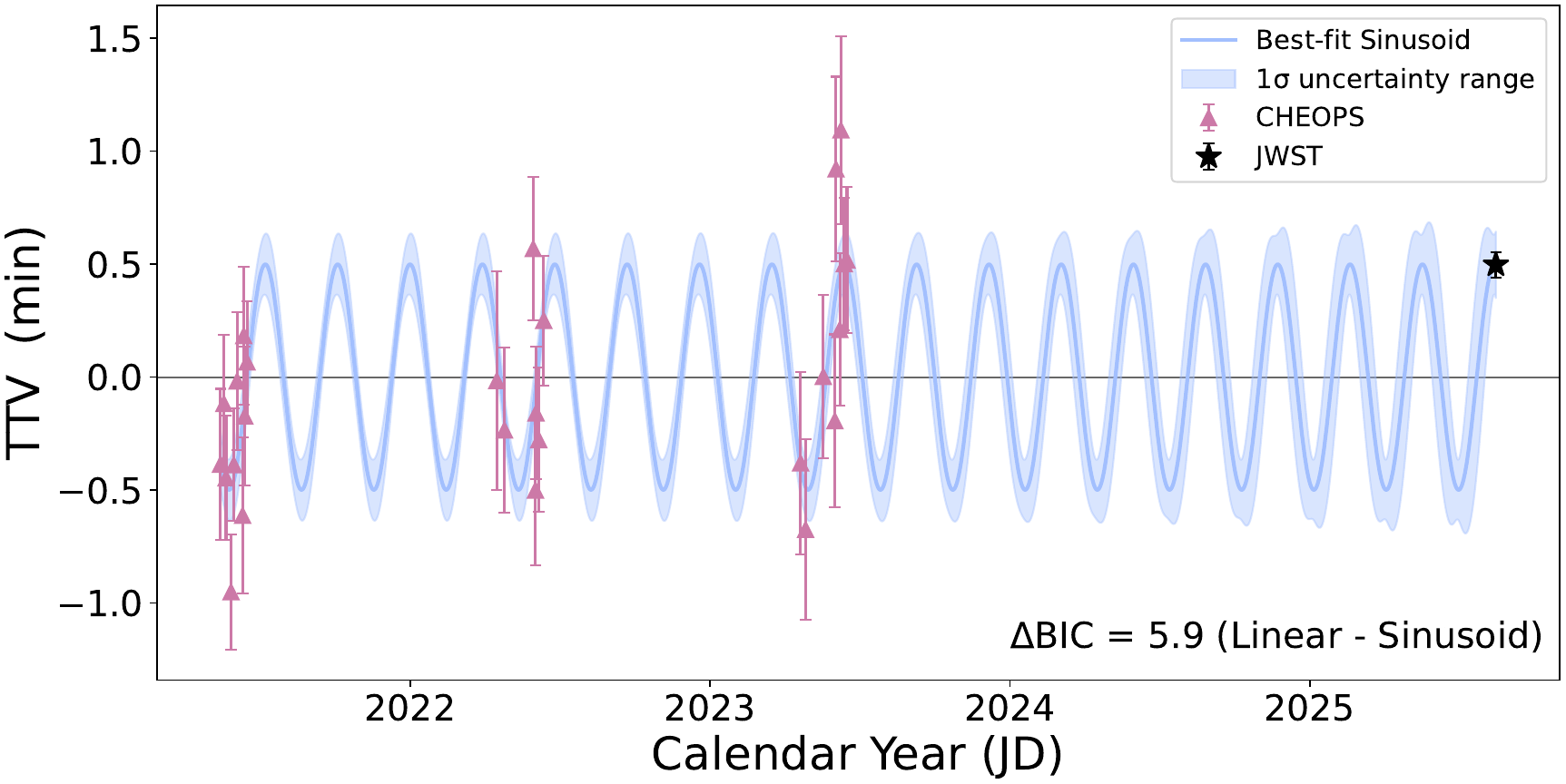}
  \caption{Transit timing variations from CHEOPS (pink triangles) and JWST (black star) and the best-fit sinusoid model, compared to a linear ephemeris. The $\Delta$BIC between a linear ephemeris and sinusoidal TTVs is 5.9, indicating strong evidence for a sinusoidal trend in the TTV data.}
  \label{fig:Sinusoid JWST}
\end{center}
\end{figure}

Typically, we would compare this apparent sinusoidal TTV signal to integer multiples of the stellar rotation period to assess the potential for biasing effects in the light curve. \cite{Wong2021} constrained the rotation period of TOI-2109 to $1.05 \pm 0.04$ days through a combination of $v$sin$i_*$ constraints from stellar spectra and the identification of periodic signals in the TESS photometry after removing the orbital phase curve signal. A second photometric variability period was also identified at 0.61 days. Because these stellar variability signals are at such short periods $\lesssim$ 1 day, it is not informative to compare the long-term sinusoidal TTV trend to integer multiples of the stellar rotation period. Instead, we discuss the possibility for this sinusoidal TTV to be related to stellar variability in Section \ref{subsubsec:stellar activity}, and highlight the low significance of the sinusoidal TTV at short oscillation periods when including more data in Section \ref{subsubsec:all 1 sub min data ttvs} and Figure \ref{fig:Sinusoid all data periods}.

\begin{figure}
\begin{center}
  \includegraphics[width=8.5cm]{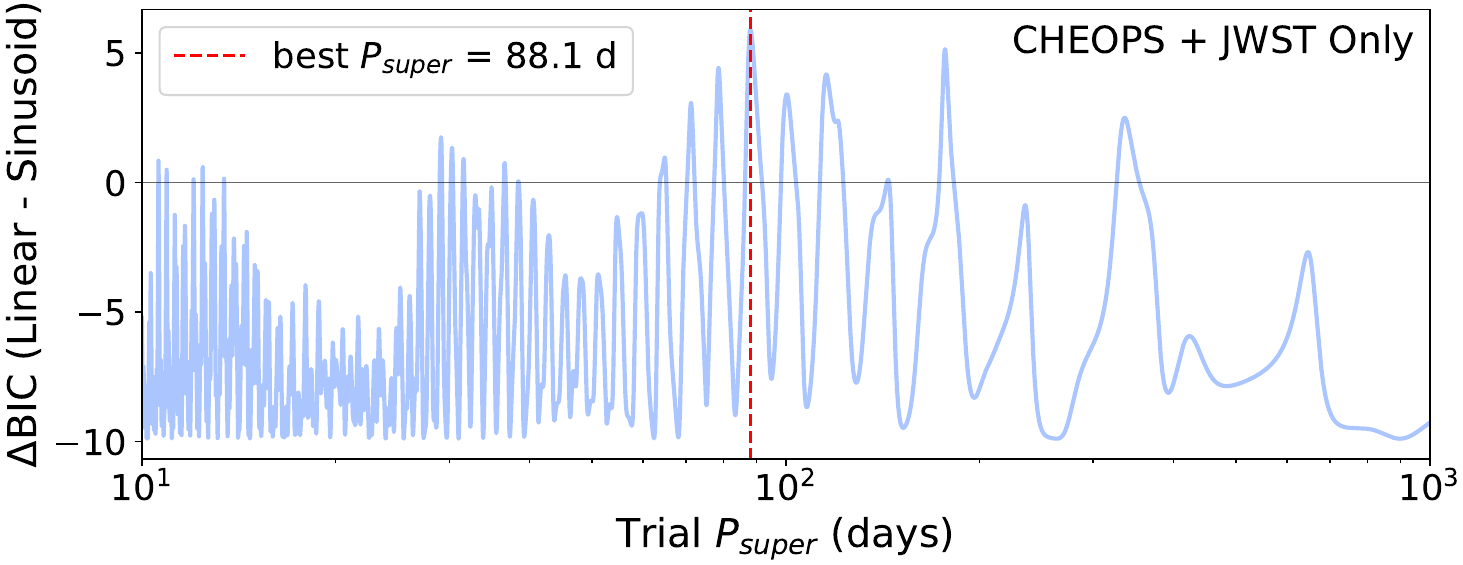}
  \caption{Comparison of significance for different sinusoidal TTV super-periods for CHEOPS + JWST data.}
  \label{fig:Sinusoid JWST periods}
\end{center}
\end{figure}

We show the sinusoidal TTVs from CHEOPS + JWST along with the best-fit model in Figure $\ref{fig:Sinusoid JWST}$, and the results of our $P_{\rm sup}$ grid search in Figure $\ref{fig:Sinusoid JWST periods}$. The multimodality of the $P_{\rm sup}$ distribution complicates any potential dynamical analysis of the TTVs as caused by a non-transiting external planetary companion. The solution is degenerate without transits of the external planet, so it is not possible to determine the exact period ratio of the planetary companion with the current data.

\subsubsection{Ground + CHEOPS + TESS + JWST Sinusoidal TTV Analysis} \label{subsubsec:all 1 sub min data ttvs}

\begin{figure*}
\begin{center}
  \includegraphics[width=\textwidth]{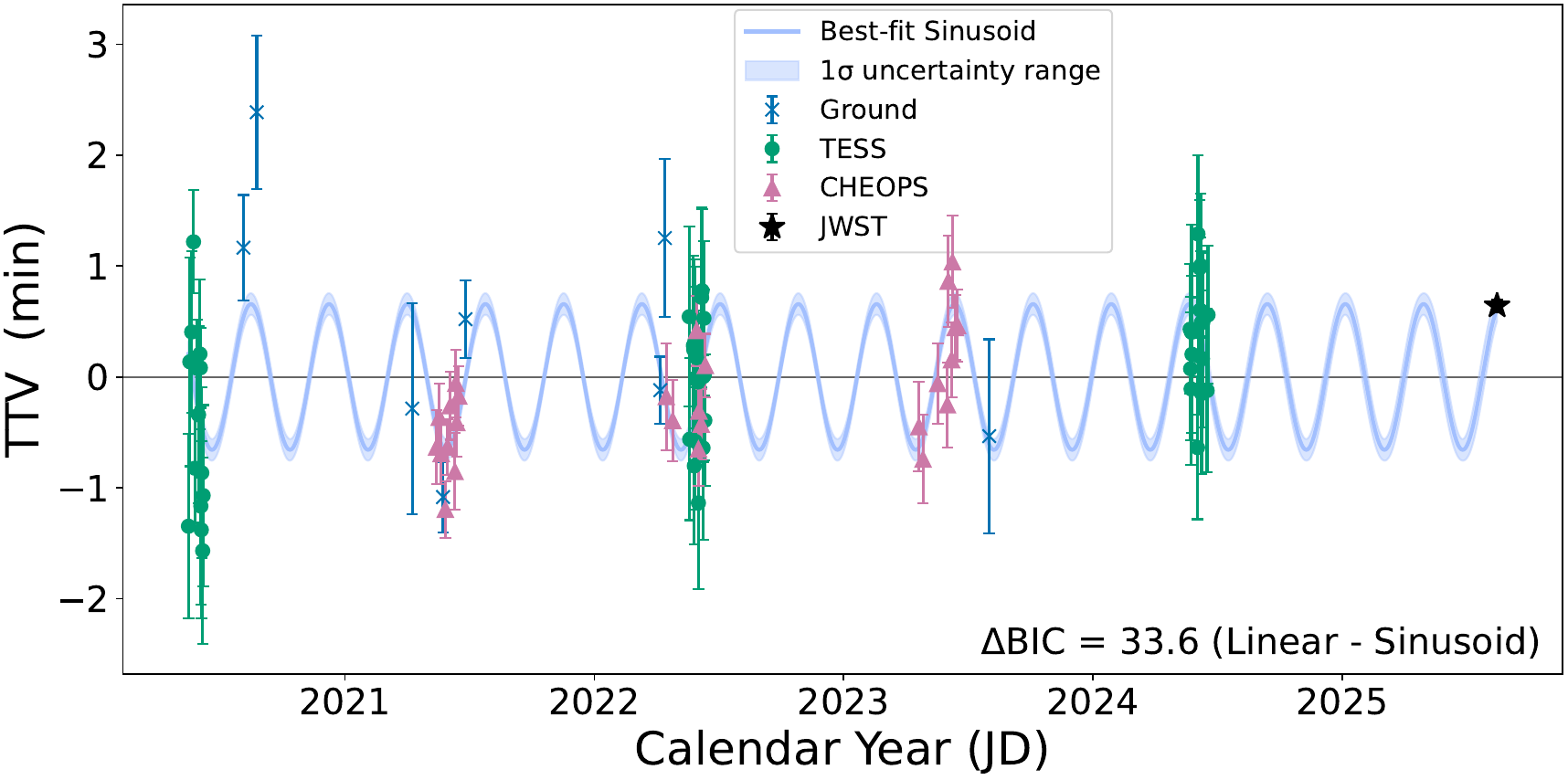}
  \caption{Transit timing variations from the highest precision data (median timing uncertainty $<$ 1 min), with the sinusoid trend fit (blue), compared to a linear ephemeris. The $\Delta$BIC between a linear ephemeris and sinusoidal TTVs is 33.6, indicating strong evidence for TTVs consistent with a nearby planetary companion.}
  \label{fig:Sinusoid all data}
\end{center}
\end{figure*}

\begin{figure}
\begin{center}
  \includegraphics[width=8.5cm]{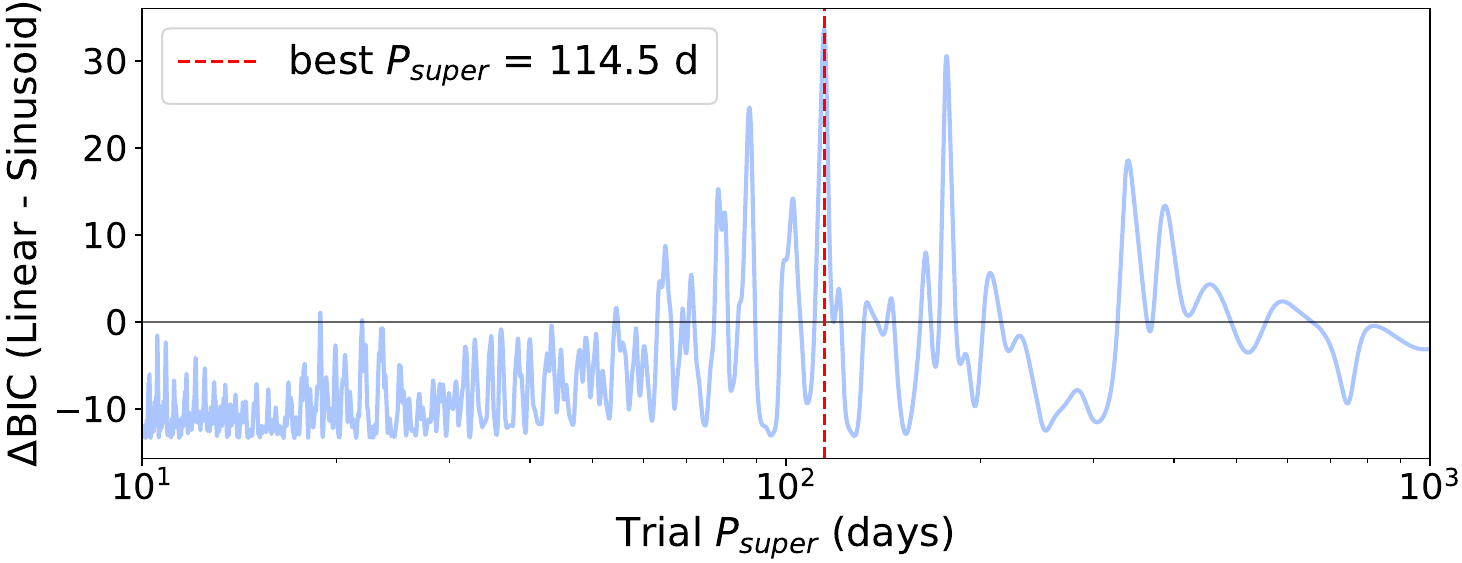}
  \caption{Comparison of significance for different sinusoidal TTV periods for all data with precision better than 1 min.}
  \label{fig:Sinusoid all data periods}
\end{center}
\end{figure}

Because the best-fit sinusoid has an amplitude of only $\sim$ 39 seconds (Figure \ref{fig:Sinusoid all data}), transit times with uncertainties much larger than this are uninformative for the fit. For this reason, JVH24 chose to use only the CHEOPS data when modeling the sinusoidal TTVs. However, several individual TESS and ground-based transits as reduced by the later work of JAM25 have timing uncertainties below one minute, and are therefore still informative given the inferred TTV amplitude. We therefore repeated the model comparison described in Section \ref{subsubsec:cheops + jwst ttvs} including these sub-minute precision TESS and ground-based transits. Including all additional high-precision data causes the evidence for the sinusoidal model to rise dramatically, with $\Delta\mathrm{BIC} = 33.6$. According to the scale of \cite{Raftery1995}, this is very strong evidence for the sinusoidal TTVs. 

As shown in Figure $\ref{fig:Sinusoid all data}$, the extreme precision of the JWST transit anchors the sinusoidal trend, which is clearly followed by the high-precision TESS, ground-based, HST, and CHEOPS timings across the full baseline. Our best-fit $P_{\rm sup}$ value of 114.52 $\pm$ 0.44 days, shown in Figure \ref{fig:Sinusoid all data periods}, is consistent within 2$\sigma$ with the original best-fit $P_{\rm sup}$ from JVH24 of 117.34 $^{+1.23}_{-1.15}$ days. If we instead include all available timings regardless of their precision other than the WASP transit time, the evidence is lower but remains decisive in favor of the sinusoidal TTV model, with $\Delta\mathrm{BIC} = 26.2$. If we also include the WASP data point in addition to all available TESS, CHEOPS, HST, JWST, and ground-based data (see Section \ref{subsec:Decay} for further details on why this point is excluded from our sinusoidal TTV analysis), the evidence decreases again, but is still firmly in the very strong range for the sinusoidal model, with  $\Delta\mathrm{BIC} = 16.7$. Therefore we can confirm that the sinusoidal TTV trend first identified by JVH24 is real. 

Given that our detected sinusoidal TTVs are consistent with the period and amplitude characterized in JVH24, it is plausible that the sinusoidal TTVs are caused by an external near-resonant non-transiting planet as described by their photodynamical analysis. As detailed in the following subsections, it is possible for us to rule out parts (but not all) of the original parameter space for a companion planet as described by JVH24, and we can also rule out many of the standard alternative non-planetary causes for the observed sinusoidal TTVs (Section \ref{subsec:Other causes of sinusoidal TTVs}). We are therefore left with a low-mass planetary companion as the most likely cause of these TTVs. However, out of an abundance of caution, we note that true confirmation of the planetary companion would determine precise constraints on the planetary properties through self consistent photo-dynamical modeling using our new transit timing measurements. %This analysis is the subject of future work beyond the scope of this paper.

\subsubsection{Limits on the Properties of a Companion Planet}

If the sinusoidal TTV is indeed produced by a planetary companion, its properties should be broadly consistent with those inferred by JVH24 given the similarity of their best-fit TTV amplitude and period compared to the values we obtain in this work. This would require a close-in (period from 1.1 to  2.4 days), large ($< 350 M_{\oplus}$) near-resonant outer companion to TOI-2109~b. Any planet that is coplanar with TOI-2109~b on these orbits would not be transiting given the inclination of TOI-2109~b \citep{Wong2021}. Given the high precision of our JWST partial phase curve (Section $\ref{subsec:obs_JST}$), we can place limits on the presence of potential co-planar outer companions based on the fact that nearby massive companions should create additional signal in the phase curve amplitude due to flux from the companion planet.

We find that for companion planets near any of the orbital resonances required to produce sinusoidal TTVs as detailed in JVH24 ($< 2.4 $ days), any planet larger than $\sim 0.1 M_{\rm Jup}$ that is coplanar with TOI-2109~b would produce a detectable phase curve variation $>$ 100 ppm, which we do not find (See Section $\ref{subsec:obs_JST}$ for more details). We determine the expected phase curve amplitudes by calculating theoretical day and night spectra using the forward modeling module from the package \texttt{Picaso} \citep{batalha_exoplanet_2019}. We approximate the planet radii from the mass and equilibrium temperature using detected exoplanets with similar properties from the Exoplanet Archive \citep{Christiansen2025}. This rules out most of the allowable parameter space for the companion planet as shown in Figure 6 of JVH24, except for low-mass planets near the 7:2 resonance with TOI-2109~b. However, an alternative configuration where the companion planet has a significant mutual inclination relative to TOI-2109~b is also possible, which might enable it to evade detection in both transit and phase curve modulation but still produce the observed sinusoidal TTVs \citep[e.g.,][]{Nabbie2025}.

Given the new high-precision JWST data which anchors the TTV curve and results in a strongly detected sinusoidal trend, a self-consistent photodynamical modeling of all the available photometric and RV data that combines all simultaneous effects for this unique system would be the most reliable way forward to determine the properties of the companion planet, or rule out the companion planet hypothesis. Dynamical effects that should be simultaneously modeled for this system include planet-planet gravitational interactions affecting the transit times and transit durations of TOI-2109~b, phase curve variations from the flux of both TOI-2109~b and any potential non-transiting companion, planet and star tidal distortion effects, stellar heterogeneities that affect photometric slopes near or in transit events, along with potential tidally-driven decay, apsidal precession, and orbital variations due to stellar quadropole moment changes. Future work on this framework will be able to confirm what region of parameter space for a potential planet c is fully consistent with these data.

\subsection{Non-planetary Causes of Sinusoidal TTVs} \label{subsec:Other causes of sinusoidal TTVs}

Given the strong detection of sinusoidal TTVs but the strict limits on any additional planetary companion due to the lack of transit and phase curve evidence, we explore potential other sources for sinusoidal TTVs in the following subsections. We do not identify any plausible non-planetary source for the sinusoidal TTVs.

\subsubsection{Apsidal Precession}

A planet on an eccentric orbit with precessing apsides can exhibit sinusoidal TTVs that potentially mimic the TTVs caused be a nearby planetary companion. Apsidal precession can be caused by many dynamical mechanisms, including but not limited to general relativistic effects \citep[e.g.,][]{Jordan2008,Pal2008}, changing stellar \citep[e.g.,][]{Miralda-Escude2002,Heyl2007} and planetary \citep[e.g.,][]{Ragozzine2009} quadrupole moments, and gravitational affects from other planets \citep[e.g.,][]{Miralda-Escude2002,Heyl2007}. We investigate the possibility that the sinusoidal TTVs of TOI-2109~b might be caused by apsidal precession, while remaining agnostic as to the cause of the precession. Using the formalism of \cite{Gimenez1995}, we can fit the TTVs caused by apsidal precession with the model:
\begin{align} \label{eq:apsidal_precession}
    t_{\rm tra}(N) &= T_0 + NP_\mathrm{s} - \frac{eP_\mathrm{a}}{\pi}\cos\omega(N),\\
    %\omega(N) &= \omega_0 + \frac{d\omega}{dN}N \nonumber \\
    P_\mathrm{s} &= P_\mathrm{a}\Big(1 - 2\pi\frac{d\omega}{dN}\Big), \nonumber
\end{align}
where $d\omega/dN$ is the precession rate, $P_\mathrm{s}$ and $P_\mathrm{a}$ are the sidereal and anomalistic periods, respectively, and $\cos\omega(N)$ corresponds to the instantaneous longitude of periastron $\omega$ value during the $N$-th transit epoch. Due to the fortunate circumstance of having our best-constrained transit epoch and best-constrained eclipse epoch from JWST both sampled one after another at the peak of our sinusoidal TTV distribution (see Figure \ref{fig:Sinusoid all data}), we can use the $e\cos\omega(N)$ value from this transit epoch $N = 10435$ to check whether apsidal precession is capable of producing the observed maximum TTV amplitude of $\sim$ 39 seconds. 

From the first JWST eclipse time and immediately following JWST transit time in Table \ref{tab:transit times}, we can calculate the secondary eclipse offset time relative to arrival at phase 0.5, which would be expected for a perfectly circular orbit. We assume a $P_\mathrm{s}$ value from a linear ephemeris fit to all sub-1 minute timing precision transit data as shown in Figure \ref{fig:Sinusoid all data} to calculate the predicted phase 0.5 eclipse time prior to our observed JWST transit time, and determine that the best-fit secondary eclipse time is offset from this value by $\Delta t_{\rm sec} = 4.19 \pm 7.59$ seconds, consistent with a circular orbit. Taking the $3\sigma$ upper limit on this secondary eclipse timing offset value (3.12 x$10^{-4}$ days) and using the formula for the secondary eclipse offset time as a function of eccentricity from equation 33 of \cite{Winn2010} ($e\cos\omega = (\Delta t_{\rm sec}\pi)/(2P_\mathrm{s})$), we get a $3\sigma$ upper limit on the $e\cos\omega(N)$ value at a transit N that corresponds to the peak of the sinusoidal TTV variation, effectively the maximum value for $\lvert{e\cos\omega(N)}\rvert$, of 0.000728. Applying this maximum $\lvert{e\cos\omega(N)}\rvert$ to Equation \ref{eq:apsidal_precession}, we find that the maximum expected deviation from a linear ephemeris in the $3\sigma$ upper limit eccentricity case due to apsidal precession is $\sim$ 13 seconds, compared to our measured sinusoidal TTV amplitude of 39 seconds. This demonstrates that no matter the cause of potential apsidal precession for TOI-2109~b, the orbital eccentricity is not large enough to explain the majority of the observed sinusoidal TTV signal.

%GR precession not possible to cause apsidal precession because the predicted $d\omega/dN$ rate is way too small to match the observed period, but not worth mentioning because apsidal precession overall doesn't work. %Since we have measured the period of sinusoidal TTVs at 114.5 days, we can constrain the rate of precession of the longitude of periastron ($d\omega/dN$) to 0.03674 rad/orbit. 

\subsubsection{The Light Travel Time Effect}

Another potential cause of sinusoidal TTVs is the light travel time effect, also known as the R\o mer effect or the Doppler effect \citep{Bouma2020}. In this scenario, TOI-2109 orbits around a common barycenter with a massive third object, which causes the transits of TOI-2109~b to arrive earlier and then later with a period and amplitude that matches the light travel time across the orbit of TOI-2109 around the common barycenter. In this case, the orbital period of the massive third object must match the orbital period of the sinusoidal TTVs, which we have constrained to  114.5 days. We calculate the minimum mass of this third object in the following way: The distance to the system barycenter that TOI-2109 must have to produce an $A_{\rm TTV} = 39$ second TTV amplitude due to light travel time is $a_1 = A_{\rm TTV}\times c = $ 0.079 AU. Then the total separation between TOI-2109 and the massive third body $a$ can be found through the barycenter mass balance equation $a_1(M_1 + M_2) = a(M_2)$ where $M_1 = 1.453 M_{\odot}$ is the mass of TOI-2109 and $M_2$ is the mass of the third body. This is then combined with Kepler's third law $a^3 = G(M_1 + M_2)P_{\rm TTV}^2/4\pi^2$ to determine that $M_2 = 0.24 M_{\odot}$ and $a = 0.55$ AU.

If there was an additional low-mass M dwarf with $M_{\star} = 0.24 M_{\odot}$ orbiting at a distance of 0.55 AU from TOI-2109, this could reproduce the observed sinusoidal TTVs from the light travel time effect, but the predicted RV semi-amplitude of such an object would be 7.5 km/s, easily detectable in the RV data \citep{Wong2021}. Therefore, the sinusoidal TTVs of TOI-2109 cannot be explained by light travel time effects. 

\subsubsection{Stellar Heterogeneities} \label{subsubsec:stellar activity}

The presence of heterogeneities on the stellar surface, such as dark starspots or bright faculae, can create artifacts on the transit shape if the planet passes in front of these regions, and these artifacts can bias the transit timing variations in a way that sometimes mimics sinusoidal TTVs \citep[e.g.,][]{Ioannidis2016}. TOI-2109 does exhibit photometric variability in the TESS data with peak to peak amplitudes of $\sim$400 ppm, compared to the transit depth of 6650 ppm, as described in \cite{Wong2021}. Manual inspection of the highest precision light curves from JWST and TESS sector 79 does not reveal any anomalies from stellar spot crossings, making it unlikely that these effects have biased the planetary transit times. Detailed simulations of transit profiles including starspot crossings found that for the most extreme cases these effects can bias the transit times by $\sim1\%$ of the transit duration \cite{Ioannidis2016}. In the case of TOI-2109~b, this would be about 1.1 minutes, but given that we detect no in-transit shape anomalies whatsoever, it is very unlikely that the sinusoidal TTV amplitude of 39 seconds is caused by undetectably small crossings of stellar spots or faculae.

\subsubsection{Applegate Mechanism}

Sinusoidal TTVs can also be caused by the Applegate Mechanism \citep{Applegate1992}. In this scenario, changes to the angular momentum of the stellar convective envelope throughout its magnetic dynamo cycle affect the stellar quadrupole and octupole moments, and these changes to the stellar shape in turn drive changes to the planetary orbit which can mimic a long-term TTV signature \citep[e.g.,][]{Watson2010}. TOI-2109 is close to the Kraft break \citep{Wang2026_kraft}, above which stars lack the deep outer convective envelopes that drive magnetic activity cycles. TOI-2109~b exhibits photometric variability that may be linked to the star's rotation period as stellar heterogeneities on an outer convective envelope rotate in and out of view, but this variability could also stem from intrinsic pulsation modes \citep{Wong2021}. If we assume that TOI-2109 does have a magnetic activity cycle, our detected sinusoidal TTV orbital period ($\sim$114.5 days) is far too short to be caused by the Applegate mechanism. The expected TTV amplitude for a given planet decreases as the magnetic dynamo modulation period decreases, and the typical expected stellar modulation periods are years to decades \citep[e.g.,][and references therein]{Watson2010}. For WASP-18~b, a comparable giant planet on an ultra-short period orbit around an F star, \cite{Watson2010} estimated that the maximum TTV amplitude caused by the Applegate mechanism is less than 10 seconds for stellar modulation periods faster than 10 years. Given the $\sim$114.5 day sinusoidal TTV period of TOI-2109~b, it is exceedingly unlikely that these TTVs are caused by modulations to the stellar dynamo. 

In principle, a sinusoidal TTV might also be caused by changes to the stellar structure that are driven by the planet itself, not necessarily related to the star's intrinsic magnetic activity cycle. Close-in planets are known to excite stellar oscillations through dynamical tides, and these oscillations can last for Gyr, but the variability periods are typically hundreds of years or more \citep{Lanza2022}. We are not aware of any proposed mechanism for star-planet interaction that would drive stellar oscillations large enough to alter the orbital period of a massive close-in planet with a variability period around 115 days.

\subsubsection{Exomoon Induced TTVs}

Another potential source of sinusoidal TTVs could be the presence of a large moon orbiting TOI-2109~b \citep[e.g.,][]{Kipping2009}. It is extremely unlikely that an exomoon around TOI-2109~b would be long-term stable, as it would face destabilization from dynamical perturbations during planet migration, tidal effects from both planet and star, and planetary magnetic field effects, all of which are expected to rapidly disrupt exomoons around hot Jupiters \citep[e.g.,][]{Heller2014,Adams2016,Trani2020,Wei2024,Bolmont2025}. 

Nevertheless, we can estimate the minimum mass of the moon required to produce the observed TTVs using Equation 1 of \cite{Simon2007}, by assuming that the moon orbits TOI-2109~b at the planetary Hill radius (maximizing the satellite's orbital distance minimizes the mass required to produce the observed TTV). Exomoons on prograde orbits are typically not stable beyond $\sim40\%$ of the planetary Hill radius, while retrograde exomoons may be stable out to $\sim90\%$ \citep{Perez-Rodrigo2026}, but we adopt the Hill radius as a conservative upper limit on the orbital separation. This results in a satellite with a mass of at least 33 $M_{\oplus}$ to produce the observed 39 second sinusoidal TTVs. Systems where a moon is inclined with respect to the orbital plane of the planet will produce the same TTVs as a system seen edge-on \citep{Teachey2021}. Planets with masses $> 30 M_{\oplus}$ almost exclusively have radii $> 2.5 R_{\oplus}$ \citep{lopez_fortney_2014}, which would be easily detectable in transit around TOI-2109 from our JWST data given the expected transit depth for such an object of 182 ppm. The radius of TOI-2109 is $\sim$ 12.3 times larger than TOI-2109~b, so despite the high impact parameter of TOI-2109~b at $b = 0.75$ \citep{Wong2021}, even a moon at the planetary Hill radius would still exhibit a transit of the star at some point during the JWST phase curve (whether the moon is between TOI-2109 and TOI-2109~b or between TOI-2109~b and the observer, depending on the inclination of the moon's orbit and timing). This means that no matter the inclination of the exomoon, any satellite massive enough to cause the observed sinusoidal TTVs should have been visible in transit with JWST. Since we do not detect any such signature, we can observationally rule out exomoons as the cause of the sinusoidal TTVs, in addition to the strong theoretical arguments for why a large exomoon should not be stable around TOI-2109~b.

\subsection{Evidence for Quadratic TTVs} \label{subsec:Decay}

TOI-2109~b has the shortest orbital period of any known gas giant planet, and it is one of the most promising candidates for the detection of tidal orbital decay. \cite{Wong2021} estimated that for stellar tidal quality factors in the range of $Q_\star' = 10^5$--$10^7$, the orbital decay rate of TOI-2109~b is $\sim$10--740\,ms\,yr$^{-1}$. There have been numerous observational campaigns that attempted to detect orbital decay since the original characterization of TOI-2109~b in \cite{Wong2021}, yielding various upper limits and marginal detections up to 3$\sigma$ \citep{Wong2021,Harre2024,Alvarado2025, Kutluay2026}.

The standard model for the transit times of a decaying orbit is a quadratic ephemeris described by:
\begin{equation} \label{eq:decay}
t_{\rm tra}(N) = T_0 + N\,P + \frac{1}{2}\frac{dP}{dN}\,N^2,
\end{equation}
where $dP/dN$ is the change in orbital period per orbit. This can be converted to the period derivative via $\dot{P} = P^{-1}\,dP/dN$ and expressed in conventional units of ms/yr. The period derivative $\dot{P}$ can then be converted to a constraint on the modified stellar tidal quality factor ($Q_\star’$) through the constant-phase-lag formulation of \cite{Goldreich1996}:
\begin{equation} \label{eq:qstar}
\dot{P} = -\frac{27\pi}{Q_\star'}\,\frac{M_{\rm p}}{M_\star}\left(\frac{R_\star}{a}\right)^5,
\end{equation}
where $M_{\rm p}/M_\star = q$ is the planet-to-star mass ratio and $a/R_\star$ is the scaled semimajor axis. When calculating $Q_\star’$, we adopt the system parameters from \cite{Wong2021} ($q = 0.00331 \pm 0.00052$ and $a/R_\star = 2.268 \pm 0.021$). We use both values from \cite{Wong2021} for consistency since our JWST observations do not provide an independent constraint on the planet-to-star mass ratio, but we note that our JWST transit fit yields an $a/R_\star$ value that is consistent within 2$\sigma$ to value of \cite{Wong2021}.

Previous efforts to measure orbital decay for TOI-2109~b have varied in their constraining power. JVH24 analyzed the same CHEOPS, TESS Sector 25 and 52, and ground-based data as this work, though without the JWST and HST data, and do not model the sinusoidal and decay TTVs simultaneously, instead fitting a sinusoid and then removing it before fitting a decay model. With this framework they measure a decay rate of $\dot{P} = -5.56 \pm 1.62$ ms/yr, placing a $2\sigma$ lower limit of $Q_\star' > 8.4 \times 10^6$. Independently, JAM25 extended the timing baseline with TESS Sector 79 and fit an orbital decay model that does not account for any sinusoidal TTVs. They measured an observed decay rate of $\dot{P} = -2.616 \pm 1.285$ ms/yr, consistent at $\sim 2 \sigma$ with a constant period, and implying a $2\sigma$ lower limit of $Q_\star' > 3.7 \times 10^7$. Their stellar tidal modeling (outside of timing data) suggested that TOI-2109 is a “young” host star with $Q_\star' > 2.3 \times 10^7$. Most recently, \cite{Kutluay2026} performed a homogeneous TTV survey of 20 hot Jupiters (without accounting for any sinusoidal trend for TOI-2109~b) and found that the TTVs of TOI-2109~b are consistent with a linear ephemeris ($\Delta\mathrm{BIC} = 1.1$) and set a $3\sigma$ lower limit of $Q_\star' > 1.23 \pm 0.20 \times 10^6$.

Given our now stronger evidence for sinusoidal TTVs (Section \ref{subsec:Sinusoidal_TTVs}), we must account for this signal in our updated search for orbital decay. Instead of the two-step process of JVH24, we adopted a self-consistent approach, where we fit the full ground-based, TESS, CHEOPS, WASP, HST, and JWST transit data set with a single joint model containing both a quadratic decay term and a sinusoidal TTV:
\begin{equation} \label{eq:joint}
t_{\rm tra}(N) = T_0 + NP + \frac{1}{2}\frac{dP}{dN}\,N^2 + A\cos\!\left[\frac{2\pi}{P_{\rm sup}}\left(NP - T_{\rm sup}\right)\right],
\end{equation}
where we optimized all six parameters simultaneously. As in Section \ref{subsec:Sinusoidal_TTVs}, we initialized our fit with a grid scan over $P_{\rm sup}$, followed by a nonlinear refinement of the full six-parameter model that is initialized from the optimal $P_{\rm sup}$ value from the grid scan and solved using the \texttt{scipy} \texttt{curve\_fit} function. We fit this joint model to all available data with a timing precision better than 1 minute, but now explicitly include the lower precision WASP transit time since its very early epoch anchors the decay trend. 

\begin{deluxetable}{ccc}
\tabletypesize{\footnotesize}
\tablecaption{Joint sinusoidal + quadratic model (Equation \ref{eq:joint}) of TTVs.\label{tab:TTV model}}
\tablehead{
  \colhead{Parameter} &
  \colhead{Prior} &
  \colhead{Posterior}
}
\startdata
$T_0$ $(BJD_{\rm TDB})$ & $\mathcal{U}( T_{0_{\rm WASP}} - 1.0, T_{0_{\rm WASP}} + 1.0)$ & -3114.3108 $\pm$ 0.0012 \\
P (d) & $\mathcal{U}(P_{\rm W21} - 0.01, P_{\rm W21} + 0.01)$ & 0.67247479 $\pm$ 0.00000028 \\
dP/dN & $\mathcal{U}(-0.01, 0.01)$ & $-1.09 \pm 0.33 \times 10^{-10}$ \\
A (min) & $\mathcal{U}(0, 100)$ & 0.94 $\pm$ 0.15 \\
$P_{\rm sup}$ (d) & $\mathcal{U}(10, 14000)$ & 177.65 $\pm$ 0.71 \\
$T_{\rm sup}$ (d) & $\mathcal{U}(0, P_{\rm sup})$ & 52.36 $\pm$ 24.95 \\
\enddata
\tablecomments{All times in $BJD_{\rm TDB}$ are in relative units of $BJD_{\rm TDB}-2457000$. $T_{0_{\rm WASP}}$ is the best-fit transit time for the WASP data from JVH24, and $P_{\rm W21}$ is the best-fit linear ephemeris orbital period from \cite{Wong2021}. Because we only fit for $T_{\rm sup}$ after performing an initial grid search to identify the most promising intialization point for $P_{\rm sup}$ (see Section \ref{subsec:Sinusoidal_TTVs}), we can remove aliases in the posterior of $T_{\rm sup}$ by limiting its prior bounds to $< P_{\rm sup}$.}
\end{deluxetable}

\begin{deluxetable}{lcccc}
\tabletypesize{\footnotesize}
\tablecaption{TTV model comparison.\label{tab:model_comparison}}
\tablehead{
  \colhead{Model} &
  \colhead{$\chi^2$} &
  \colhead{$k$} &
  \colhead{BIC} &
  \colhead{$\Delta$BIC}
}
\startdata
Decay $+$ sinusoid  & 124.46 & 6 & 152.60 &  0.00 \\
Linear $+$ sinusoid & 135.89 & 5 & 159.35 &  6.75 \\
Linear only         & 166.35 & 2 & 175.73 & 23.13 \\
Decay only          & 164.28 & 3 & 178.35 & 25.75 \\
\enddata
\tablecomments{$k$ is the number of free parameters, and $\Delta$BIC is measured
relative to the preferred (lowest-BIC) decay $+$ sinusoid model. We perform these model comparisons by successively removing terms from Equation \ref{eq:joint} (linear + sinusoid has no cosine term, and decay only has no dP/dN term), and then optimizing the model based on the timing data in Table \ref{tab:transit times}. }
\end{deluxetable}

\begin{figure*}
\begin{center}
  \includegraphics[width=18cm]{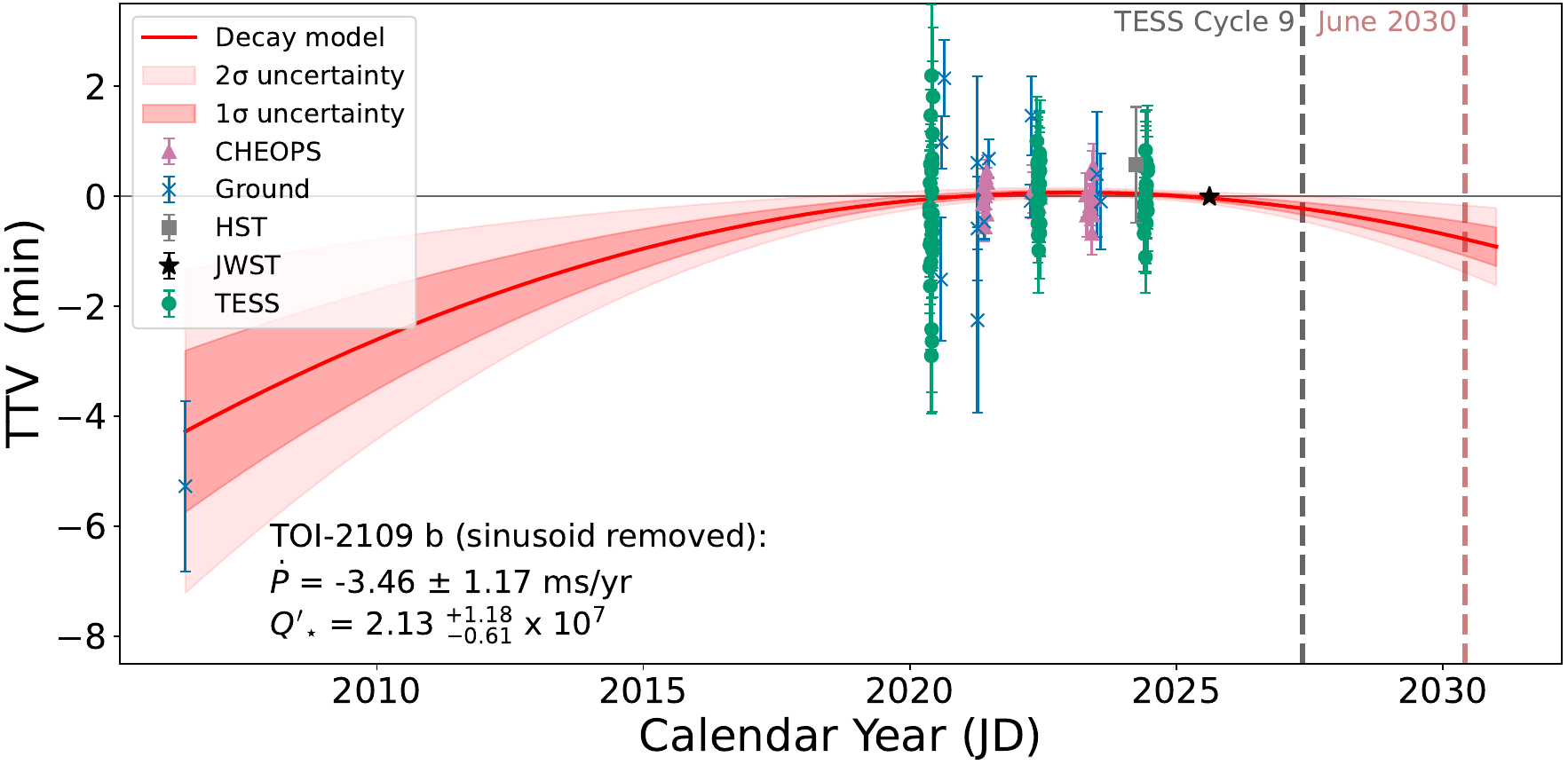}
  \caption{Transit times of TOI-2109~b, including previously published ground-based data \citep[blue crosses, ][]{Alvarado2025}, space-based data \citep[green circles for TESS, purple triangles for CHEOPS, ][]{Harre2024}, and our new HST (gray square) and JWST (black star) transit observations, with the best-fit sinusoidal TTV (Section \ref{subsec:Sinusoidal_TTVs}) removed from the data. The best-fit orbital decay model is shown in red, with the 1-2 $\sigma$ uncertainty ranges highlighted. The next TESS observations in Cycle 9 are marked with a black dashed line. If the current decay model is accurate and a transit observation with 10 second timing precision was obtained in June 2030 (the red dashed line), the decay detection would improve to $\sim$5$\sigma$.}
  \label{fig:Decay Plot}
\end{center}
\end{figure*}

To determine whether the data favor orbital decay in addition to the sinusoidal TTV model, we compared four possible models across the full data set: a linear ephemeris (parameters $T_0$ and $P$), a decay-only quadratic ephemeris (Equation \ref{eq:decay}), a linear-plus-sinusoid model (Equation \ref{eq:sinusoid}), and the joint decay-plus-sinusoid model (Equation \ref{eq:joint}). The resulting $\Delta$BIC values are reported in Table \ref{tab:model_comparison}. Models including a sinusoidal component are strongly preferred over both the linear and decay-only models, which is consistent with the results of Section~\ref{subsec:Sinusoidal_TTVs}. In addition, when we compare the two sinusoidal models to each other, the inclusion of the decay term is also strongly preferred relative to the model with no decay, by $\Delta\mathrm{BIC} = 6.75$, indicating that TOI-2109~b exhibits quadratic TTVs in addition to sinusoidal TTVs. 

In principle, secondary eclipse timing data can also serve as a useful tool for independently constraining the quadratic TTVs. The secondary eclipse of TOI-2109~b was detected in TESS sector 25, though with a poor timing precision of $\sim$90 seconds \citep{Wong2021}. Our HST eclipse timing precision is also $>$ 1 minute, so the only high-precision ($<$ 1 minute) eclipse timing data we have is a single datum from JWST (Table \ref{tab:transit times}). These eclipse times are not currently useful for independently constraining the quadratic TTV trend because they span an observational baseline of $\sim$5.3 years from May 2020 to August 2025, over which the expected quadratic TTV signal is $\sim$8 seconds (Figure \ref{fig:Decay Plot}), far below the precision of this data. However, future high-precision eclipse timing measurements at later epochs that significantly extend the observational baseline may be useful for adding independent evidence for the quadratic TTV trend. 

We caution that these quadratic TTVs may not necessarily be caused by orbital decay (see Section \ref{subsec:LOS Acceleration}), and that our detection of quadratic TTVs consistent with potential orbital decay depends entirely on the accuracy of the WASP transit time. If we remove the WASP data point from our joint sinusoidal + decay TTV fitting procedure, the evidence for the decay trend drops to $\Delta\mathrm{BIC} = 1.2$, which is statistically insignificant. Since this one datum has outsized impact on the final $\Delta\mathrm{BIC}$, it is possible that this datapoint is anomalous \citep[e.g.,][]{Adams2024,Jackson2023,Jackson2026}. Independent confirmation of the quadratic TTV trend at later epochs would be the most powerful method to confirm the validity of this signal. Another JWST-quality timing measurement ($<$ 5 seconds precision) in 2027 or later would improve the evidence to strong, or additional TESS observations with similar precision to the Sector 79 data ($\sim40$ seconds per transit) in Cycle 9 and then again in mid-2030 (during a hypothetical TESS Cycle 11) could achieve the same effect (Figure \ref{fig:Decay Plot}). 

From our best-fit joint decay+sinusoid model, we measure a decay rate of $-3.46 \pm 1.17$ ms/yr, which corresponds to a modified stellar tidal quality factor of $Q_\star' = 2.13^{+1.18}_{-0.61} \times 10^7$. This value is consistent with the upper end of the range expected for F stars \citep{Ogilvie2007, Lanza2011}, and is broadly consistent with the constraints of JVH24 and JAM25. If this decay rate and $Q_\star’$ constraint can be confirmed, it implies a tidal decay time of $P/\dot{P} = 16.8^{+8.0}_{-4.1}$ Myr for TOI-2109~b.

We also note that the presence of a nearby planetary companion would also likely influence any orbital decay of TOI-2109~b. As described in JVH24 and JAM25, gravitational perturbations from the companion could potentially excite the eccentricity of TOI-2109~b, which would lead to apsidal precession and affect decay rates, while angular momentum exchange between the planets could directly modify the orbital evolution of TOI-2109~b. Any future claimed detection of orbital decay for TOI-2109~b must first characterize the properties of the companion planet to determine the interwoven effects of multi-planet dynamical perturbation and tidal orbital decay.

\subsection{Line of Sight Acceleration} \label{subsec:LOS Acceleration}

In Section \ref{subsec:Other causes of sinusoidal TTVs}, we showed that the maximum possible TTV signal due to apsidal precession is less than 5 seconds given our constraint on the orbital eccentricity of TOI-2109~b from the JWST secondary eclipse timing offset. Since our quadratic TTV trend has an amplitude of $\sim$5 minutes, we can already verify that the quadratic TTVs are not caused by apsidal precession. However, another possible non-decay related cause of quadratic TTV trends is line-of-sight acceleration. 

If the TOI-2109 system is accelerating towards us along our line of sight \citep[e.g.,][]{Bouma2020}, this would also cause the transit times at later epochs to arrive earlier than expected. If this effect causes the observed period derivative for TOI-2109~b, the line of sight acceleration would manifest as a linear trend in the RV data, with a magnitude of $\dot{v}_r = c\dot{P}/P = -0.072 \pm {0.022}$~m~s$^{-1}$d$^{-1}$. Given the $\sim$240 day observational baseline of the available RV data in \cite{Wong2021}, this acceleration would produce a -17 m/s linear trend, which is impossible to detect given the $> 400$ m/s median precision on the RV data.

We can convert this line of sight acceleration into a constraint on the companion mass required to produce the observed trend through $\dot{v}_r = GM_{\rm comp}/d_{\rm comp}^2$. The high-resolution imaging of \cite{Wong2021} found TOI-2109 to be a single star with no companion brighter than 5–9 mag below the target star’s brightness from the diffraction limit ($\sim$20 mas) out to 1.2 arcseconds, which at the distance of TOI-2109 (262 pc) corresponds to physical separations from 5 to 314 au. For a companion beyond the outer speckle imaging limit of 314 au, the companion mass would have to be $> 10 M_{\odot}$ and therefore be easily detectable by Gaia. The Gaia DR3 Renormalized Unit Weight Error (RUWE) score for TOI-2109 is 1.024 \citep{GaiaCollaboration2023}, indicating that the astrometric measurements are well fit by a single star solution. However, for a companion at the 5 au inner sensitivity edge of the speckle imaging, the corresponding companion mass is 3.7 $M_{\rm Jup}$. The RV semiamplitude of this object would be $\sim$40 m/s with an orbital period of 9.3 years, which is undetectable for TOI-2109.

We can also estimate the Gaia RUWE score of this configuration using Equation 2 of \cite{Belokurov2020}, which relates the RUWE score to the on-sky angular perturbation of a single source and the mean value of the along-scan source centroiding error as a function the source's Gaia magnitude, which is read off the blue curve in Figure 9 of \cite{Lindegren2018}. See \cite{Belokurov2020} for further details. We calculate the maximum on-sky angular perturbation ($\alpha_{\star}$) for a 3.7 $M_{\rm Jup}$ companion at 5 au from $\alpha_{\star} = (M_{\rm comp}/M_{\star})\times(a_{\rm comp}/d)$, and then take the fraction of this that would be probed by Gaia DR3's 34 month observation period \citep{GaiaCollaboration2023} to obtain $\alpha_{\star, \rm Gaia}$ = 21.5 $\mu$as, and RUWE = 1.003. This is well below the threshold of 1.4 for indication of a secondary source \citep{GaiaCollaboration2023}, and consistent with the observed RUWE of 1.024, though there are many sources of excess astrometric jitter and this does not imply evidence in favor of a distant massive companion. In general, any giant planet companions within a few au or brown dwarf companions within a few tens of au could plausibly evade detection in the Gaia astrometric data, RV data, and speckle imaging. TOI-2109 is also not included in the Gaia Hipparcos catalog of accelerations \citep{Brandt2018}, so we do not have an independent constraint on the stellar acceleration. We therefore cannot conclusively rule out line of sight acceleration as a possible cause of the observed quadratic TTVs.

\section{Conclusions} \label{sec:Conclusions}

TOI-2109~b is a super-Jupiter with the shortest orbital period of any known giant planet, making it one of the most promising targets for the detection of tidal orbital deacy. Following the tentaive detection of sinusoidal TTVs identified in JVH24, TOI-2109~b is also a candidate member of the rare class of close-in giant planets with nearby planetary companions \citep[e.g.,][]{Hord2021,Wu2023,Sha2026}. Confirming the signal of orbital decay would determine the efficiency of tidal dissipation ($Q_\star'$) in TOI-2109 and constrain the remaining lifetime of the planet, while firmly detecting a companion would identify TOI-2109~b as one of a small group of ultra-hot Jupiters that likely reached their present orbits through a dynamically quiescent migration pathway (e.g., not high-eccentricity migration).

To update the dynamical constraints on this system, we combined the full set of archival ground-based, TESS, and CHEOPS photometry used in previous analyses of TOI-2109~b \citep{Wong2021, Harre2024, Alvarado2025} with new, high-precision transit and eclipse observations from JWST and HST. The JWST transit provides a timing precision of $\sim$3 seconds that anchors the TTV baseline. We fit the TTVs with linear, quadratic, and sinusoidal models and compare the evidence for each signal via the $\Delta$BIC values. 
%We assessed potential non-planetary explanations for the sinusoidal TTV signal but did not identifying any promising sources. 

We find very strong evidence for sinusoidal TTVs over a linear ephemeris ($\Delta$BIC = 33.6), confirming the signal first reported by JVH24 that is consistent with a nearby, non-transiting, near-resonant planetary companion. We rule out the standard non-planetary explanations for these sinusoidal TTVs, including apsidal precession, light travel time effects, stellar heterogeneities, the Applegate mechanism, and exomoons. Yet, our high-precision JWST phase curve also rules out much of the previously allowed companion planet parameter space, except in the case of lower mass planets at further resonances, or a mutually inclined companion. We therefore claim that a planetary companion is the most likely cause of the sinusoidal TTV signal, but further detailed modeling is required to better constrain the properties of this planet, confirm that it is consistent with all data sources, and account for potentially biasing effects such as orbital decay and apsidal precession.

When we account for the sinusoidal TTVs in a joint model of sinusoidal TTVs and orbital decay, we also find strong evidence in favor of a quadratic TTV trend ($\Delta$BIC = 6.75). This is consistent with an orbital decay rate of $\dot{P} = -3.46 \pm 1.17$~ms\,yr$^{-1}$ and a modified stellar tidal quality factor of  $Q_\star' = 2.13^{+1.18}_{-0.61} \times 10^{7}$. This is consistent with the upper end of the range expected for F stars \citep{Ogilvie2007, Lanza2011} and consistent with the previous decay rate constraints of JVH24 and JAM25. We caution that the decay signal detection depends entirely on the earliest, lower-precision WASP transit time, and that we cannot yet rule out line of sight acceleration as the source of the quadratic trend. However, as highlighted in Figure \ref{fig:Decay Plot}, additional high-precision ($<$ 10 second) observations during mid 2027 or beyond could enable confident $> 5 \sigma$ detections of the quadratic TTV trend that do not depend entirely on the WASP data. 

If the sources of the quadratic and sinusoidal TTV trends can be confirmed with future dynamical modeling and additional observations, TOI-2109~b would be just the second exoplanet with a firmly detected decaying orbit, and an extremely rare ultra-hot Jupiter with a nearby planetary companion. A single additional JWST-quality timing measurement, or continued high-precision TESS monitoring through Cycle 9 and beyond, would be sufficient to raise the evidence for the quadratic TTV to a very strong level. However, ruling out line of sight acceleration as a possible cause of the quadratic TTV trend will be challenging, especially with RV observations given the fast rotation of TOI-2109.

\begin{acknowledgments}

%We thank the anonymous referee for their constructive comments and suggestions, which helped improve the quality of the manuscript.

This work is based in part on observations made with the NASA/ESA/CSA James Webb Space Telescope. The data were obtained from the Mikulski Archive for Space Telescopes at the Space Telescope Science Institute, which is operated by the Association of Universities for Research in Astronomy, Inc., under NASA contract NAS 5-03127 for JWST. These observations are associated with program \#8877. Support for program \#8877 was provided by NASA through a grant from the Space Telescope Science Institute, which is operated by the Association of Universities for Research in Astronomy, Inc., under NASA contract NAS 5-03127. %The specific observations analyzed can be accessed via DOI: \textcolor{red}{insert DOI link}. 

This work is based in part on observations made with the NASA/ESA Hubble Space Telescope, obtained from the Data Archive at the Space Telescope Science Institute. These observations are associated with program \#17543, PI G. Fu. Support for program \#17543 was provided by NASA through a grant from the Space Telescope Science Institute, which is operated by the Association of Universities for Research in Astronomy, Inc., under NASA contract NAS5-26555.

This research has made use of the NASA Exoplanet Archive \citep{Christiansen2025} and the Exoplanet Follow-up Observation Program website, which are operated by the California Institute of Technology, under contract with the National Aeronautics and Space Administration under the Exoplanet Exploration Program. This paper includes data collected by the TESS mission that are publicly available from the Mikulski Archive for Space Telescopes. We acknowledge the use of public TESS data from pipelines at the TESS Science Office and at the TESS Science Processing Operations Center. Funding for the TESS mission is provided by NASA’s Science Mission Directorate.

This work made use of Anthropic Claude Opus 4.8, accessed in August and September 2026, for assistance with writing python code that was used for the generation of figures 2, 4, and 5, for converting python code formulae into latex equations, for converting csv files into latex format for our tables, and for suggestions on how to reduce our abstract draft to less than 250 words. Every line of output from this LLM was reviewed and verified before implementation, and the authors take full responsibility for the accuracy and integrity of the work.

\end{acknowledgments}

\begin{contribution}

Coauthor contributions are as follows: M.G.M. led the modeling and interpretation of this study, contributed to data analysis, and wrote the manuscript. J.I. and N.L.W. contributed to project conceptualization, project administration, data analysis for JWST and HST, data and model interpretation, and writing. S.V. contributed to project conceptualization, independent checks on the TTV modeling, interpretation, and detailed manuscript review. I.W. provided independent TESS data reductions and orbital decay calculations. All authors provided detailed comments and conversations that greatly improved the quality of the manuscript.

\end{contribution}

%% Use only facilities from the official AAS Journal list;
%% Each keyword is check against their list during copy editing.  
%% Individual instruments can be provided in parentheses after the keyword, but they are not verified.
%% https://journals.aas.org/facility-keywords/
\facilities{HST(WFC3), JWST(NIRSpec), TESS, CHEOPS}

\software{\texttt{exoplanet} \citep{exoplanet:joss,
exoplanet:zenodo} and its dependencies \citep{exoplanet:foremanmackey17,
exoplanet:foremanmackey18, exoplanet:agol20, exoplanet:arviz,
exoplanet:astropy13, exoplanet:astropy18, exoplanet:luger18, exoplanet:pymc3,
exoplanet:theano}
\texttt{astropy} \citep{astropy},
\texttt{scipy} \citep{scipy},
\texttt{numpy} \citep{numpy},
\texttt{matplotlib} \citep{matplotlib},
\texttt{emcee} \citep{emcee},
\texttt{corner} \citep{corner}, and
\texttt{lightkurve} \citep{lightkurve}}

%% Appendix material should be preceded with a single \appendix command.
%% There should be a \section command for each appendix. Mark appendix
%% subsections with the same markup you use in the main body of the paper.
%%
%% Each Appendix (indicated with \section) will be lettered A, B, C, etc.
%% The equation counter will reset when it encounters the \appendix
%% command and will number appendix equations (A1), (A2), etc. The
%% Figure and Table counter will not reset.

%\appendix

%\section{Extracted Transit Times}

%Fill in some text here if we decide to put in an appendix.

%% For this sample we use BibTeX plus aasjournalv7.bst to generate the
%% the bibliography. The sample7.bib file was populated from ADS. To
%% get the citations to show in the compiled file do the following:
%%
%% pdflatex sample7.tex
%% bibtext sample7
%% pdflatex sample7.tex
%% pdflatex sample7.tex

\bibliography{every_citation_ever.bib}{}
\bibliographystyle{aasjournalv7}

%% This command is needed to show the entire author+affiliation list when
%% the collaboration and author truncation commands are used.  It has to
%% go at the end of the manuscript.
%\allauthors

%% Include this line if you are using the \added, \replaced, \deleted
%% commands to see a summary list of all changes at the end of the article.
%\listofchanges

\end{document}